\documentclass[12pt]{article}

\usepackage{epsfig,latexsym}
\usepackage{amsmath}

\newcommand{\qed}{\hfill $\Box$}

\newtheorem{theo}{Theorem}
\newtheorem{lemm}{Lemma}

\newtheorem{obse}{Observation}

\title{Independent sets in ($P_4+P_4$,Triangle)-free graphs}

\author{Raffaele Mosca}

\begin{document}

\maketitle

\begin{center}
{\footnotesize
Dipartimento di Economia, Universit\'a degli Studi "G. d'Annunzio", Pescara 65127, Italy.\\
\texttt{r.mosca@unich.it} }
\end{center}




\begin{abstract}

The Maximum Weight Independent Set Problem (WIS) is a well-known NP-hard problem. A popular way to study WIS is to detect graph classes for which WIS can be solved in polynomial time, with particular reference to hereditary graph classes, i.e., defined by a hereditary graph property or equivalently by forbidding one or more induced subgraphs. Given two graphs $G$ and $H$, $G+H$ denotes the disjoint union of $G$ and $H$.


This manuscript shows that (i) WIS can be solved for ($P_4+P_4$, Triangle)-free graphs in polynomial time, where a $P_4$ is an induced path of four vertices and a Triangle is a cycle of three vertices, and that in particular it turns out that (ii) for every ($P_4+P_4$, Triangle)-free graph $G$ there is a family ${\cal S}$ of subsets of $V(G)$ inducing (complete) bipartite subgraphs of $G$, which contains polynomially many members and can be computed in polynomial time, such that every maximal independent set of $G$ is contained in some member of ${\cal S}$. These results seem to be harmonic with respect to other polynomial results for WIS on certain [subclasses of] $S_{i,j,k}$-free graphs and to other structure results on [subclasses of] Triangle-free graphs.




\end{abstract}

\section{Introduction}

For any missing notation or reference let us refer to \cite{BraLeSpi1999}.

For any graph $G$, let $V(G)$ and $E(G)$ denote respectively the vertex-set and the edge-set of $G$. Let $G$ be a graph. For any subset $U \subseteq V(G)$, let $G[U]$ denote the subgraph of $G$ induced by $U$. For any vertex-set $U \subseteq V(G)$, let $N(U) = \{v \in V(G) \setminus U:$ $v$ is adjacent to some $u \in U\}$ be the {\em neighborhood of U in G}. In particular: if $U = \{u_1,\ldots,u_k\}$, then let us simply write $N(u_1,\ldots,u_k)$ instead of $N(\{u_1,\ldots,u_k\})$; for any vertex-set $W \subseteq V(G)$, with $U \cap W = \emptyset$, let us write $N_W(U) = N(U) \cap W$. For any vertex-set $U \subseteq V(G)$, let us say that $A(U) = V(G) \setminus (U \cup N(U))$ is the {\em anti-neighborhood of U in G}. For any vertex $v \in V(G)$ and for any subset $U \subset V(G)$ (with $v \not \in U$), let us say that: $v$ $contacts$ $U$ if $v$ is adjacent to some vertex of $U$; $v$ is $partial$ to $U$ if $v$ contacts $U$ but is non-adjacent to some vertex of $U$; $v$ is $universal$ to $U$ if $v$ is adjacent to all vertices of $U$.

A {\em component of $G$} is a maximal connected subgraph of $G$. A component of $G$ is $trivial$ if it is a singleton, and $nontrivial$ otherwise. A {\em component-set of $G$} is the vertex set of a component of $G$. A component-set of $G$ is $trivial$ if it is a singleton, and $nontrivial$ otherwise. A $clique$ of $G$ is a set of pairwise adjacent vertices of $G$. An {\em independent set} (or a {\em stable set}) of $G$ is a subset of pairwise nonadjacent vertices of $G$. An independent set of $G$ is {\em maximal} if it is not properly contained in another independent set of $G$.



A graph $G$ is {\em $H$-free}, for a given graph $H$, if $G$ contains no induced subgraph isomorphic to $H$; in particular $H$ is called a forbidden induced subgraph of $G$. A graph class is {\em hereditary} if it is defined by a hereditary graph property or equivalently by forbidding a family of induced subgraphs. Given two graphs $G$ and $F$, $G + F$ denotes the disjoint union of $G$ and $F$; in particular $lG = G + G + \ldots + G$ denotes the disjoint union of $l$ copies of $G$.

A graph $G$ is $bipartite$ if $V(G)$ admits a partition $\{A,B\}$ such that $A$ and $B$ are independent sets of $G$, i.e., such that $E(G) \subseteq A \times B$; in particular $G$ is {\em complete bipartite} if $E(G) = A \times B$.



The following specific graphs are mentioned later. A $P_k$ has vertices $v_1,v_2,\ldots,v_k$ and edges $v_jv_{j+1}$ for $1 \le j < k$. A $C_k$ has vertices $v_1,v_2,\ldots,v_k$ and edges $v_jv_{j+1}$
for $1 \le j < k$ and $v_kv_1$. A $K_n$ is a complete graph of $n$ vertices.
A $Claw$ has vertices $a,b,c,d$, and edges $ab,ac,ad$. A $Fork$ has vertices
$a,b,c,d,e$, and edges $ab,ac,ad,de$ (then a Fork contains a Claw as an induced subgraph). A $S_{i,j,k}$ is the graph obtained from a Claw by subdividing respectively its edges into $i$, $j$, $k$ edges 
(e.g., $S_{0,1,2}$ is $P_4$, $S_{1,1,1}$ is Claw). \\





The {\em Maximum Weight Independent Set Problem} ({\em WIS}) is the
following: Given a graph $G$ and a weight function $w$ on
$V(G)$, determine an independent set of $G$ of maximum weight, where the weight of an independent set $I$ is given by the sum of $w(v)$ for $v \in I$.
Let $\alpha_w(G)$ denote the maximum weight of any independent set of $G$.
The WIS problem reduces to the {\em IS} problem if all vertices
$v$ have the same weight $w(v) = 1$.

The WIS problem is NP-hard \cite{GareyJohn1979}. It remains NP-hard under various restrictions, such as e.g. Triangle-free graphs \cite{Polja1974} and more generally graphs with no induced cycle of given length \cite{Murphy1992,Polja1974}, cubic graphs \cite{GareyJohn1977} and more generally $k$-regular graphs \cite{FriHedJac}, planar graphs \cite{GareyJohnStock1976}. It can be solved in polynomial time for various graph classes, such as e.g. $P_4$-free graphs \cite{CorLerSte2004}, bipartite graphs \cite{AhuMagOrl1993,DesHak1970,GroLovSch1988}
and more generally perfect graphs \cite{GroLovSch1984}, Claw-free graphs \cite{FaeOriSta2011,Minty1980,NakTam2001,NobSas2011,Sbihi1980} and
more generally Fork-free graphs \cite{Aleks2004/1,LozMil}, 
$2K_2$-free graphs \cite{Farber1989} and more generally $lK_2$-free graphs for any constant $l$ (by combining an algorithm
generating all maximal independent sets of a graph \cite{TsuIdeAriShi1977} and a polynomial upper bound on the number of
maximal independent sets in $lK_2$-free graphs \cite{Aleks1991,Farber1993,Prisner1995}), $K_2+$Claw-free graphs \cite{LozMos2005}, $2P_3$-free graphs \cite{2P3}, and more generally $lP_3$-free graphs for any constant $l$, and $lClaw$-free graphs for any constant $l$ \cite{BraMoslClaw}; then recently, after many attempts, for $P_5$-free graphs \cite{P5} and more generally for $P_6$-free graphs \cite{P6}.









Let us report the following result due to Alekseev \cite{Aleks1983,Aleks2004/2}.



\begin{theo}\label{theo:Alekseev}{\bf \cite{Aleks1983}}
Let $X$ be a class of graphs defined by a finite set $M$ of forbidden induced subgraphs. If $M$ contains no graph every connected component of which is $S_{i,j,k}$ for some indices $i,j,k$,
then the (W)IS problem is {\em NP}-hard in the class $X$.
\end{theo}



Theorem \ref{theo:Alekseev} implies that (unless P = NP) for any graph $F$, if WIS can be solved for $F$-free graphs in polynomial time,
then each connected component of $F$ is $S_{i,j,k}$ for some indices $i,j,k$.
Then Lozin \cite{Lozin2017} conjectured that WIS can be solved in polynomial time for $S_{i,j,k}$-graphs for any fixed indices $i,j,k$.
The above allows one to focus on possible open problems, i.e., on possible graph classes for which WIS may be solved in polynomial time.

This manuscript shows that (i) WIS can be solved for ($P_4+P_4$, Triangle)-free graphs in polynomial time, and that in particular it turns out that (ii) for every ($P_4+P_4$, Triangle)-free graph $G$ there is a family ${\cal S}$ of subsets of $V(G)$ inducing (complete) bipartite subgraphs of $G$, which contains polynomially many members and can be computed in polynomial time, such that every maximal independent set of $G$ is contained in some member of ${\cal S}$.

The class of $P_4+P_4$-free graphs has been considered since, according to the above mentioned polynomial results and to possibly forthcoming similar polynomial results, it may be one of the next boundary graph classes for which the complexity of WIS is an open problem.


The class of Triangle-free has been considered in the context of similar previous manuscripts on other subclasses of Triangle-free graphs, namely, on ($P_7$,Triangle)-free graphs \cite{BraMosP7K3} $-$ see \cite{MafPas2016} for an extension of this result $-$ and more generally on ($S_{1,2,4}$,Triangle)-free graphs \cite{BraMosS124K3}.

However let us observe that Lozin's conjecture is open also for those $S_{i,j,k}$-graphs for any fixed indices $i,j,k$ which in addition are Triangle-free $-$ recalling that WIS remains NP-hard for Triangle-free graphs $-$ that is for restricted and more studied graph classes. Let us mention just a recent strong result due to Pilipczuk et al. \cite{MSTT2020} stating that graphs containing no Theta [a $Theta$ is a graph made of three internally vertex-disjoint chordless paths $P1 = a . . . b$, $P2 = a . . . b$, $P3 = a . . . b$ of length at least 2 and such that no edges exist between the paths except the three edges incident to $a$ and the three edges incident to $b$], no Triangle, and no $S_{i,j,k}$ as induced subgraphs for any fixed indices $i,j,k$ have bounded treewidht, which implies that a large number of NP-hard problems can be solved in polynomial time for such graphs, in particular the WIS problem.

\section{Independent sets in ($P_4+P_4$, Triangle)-free graphs}

In this section let us show that WIS can be solved for ($P_4+P_4$, Triangle)-free graphs in polynomial time.

First let us introduce two general observations: they are easy to prove and are the basis of the approach $-$ in other contexts called {\em anti-neighborhood approach} $-$ which will be used later.

\begin{obse}\label{obse:1}
For any graph $G$, $\alpha_w(G) = max \{\alpha_w(G[V(G) \setminus N(v)]): v \in V(G)\}$; then for any $v \in V(G)$, $\alpha_w(G) = max \{\alpha_w(G[V(G) \setminus N(v)]), \alpha_w(G[V(G) \setminus \{v\}]) \}$  \qed
\end{obse}


\begin{obse}\label{obse:2}
For any graph $G$ and for any order $v_1,v_2,\ldots,v_n$ of the vertices of $G$, $\alpha_w(G) = max \{\alpha_w(G[V(G) \setminus N(v_1)]),
\alpha_w(G[(V(G) \setminus \{v_1\})\setminus N(v_2)]), \ldots, \alpha_w(G[(V(G) \setminus \{v_1,\ldots,v_{n-1}\})\setminus N(v_n)])\}$.  \qed
\end{obse}

For any induced $P_4$ of any ($P_4+P_4$, Triangle)-free $G$, say $P$, of vertex set $V(P) = \{a,b,c,d\}$ and edge set $E(P) = \{ab,bc,cd\}$, one has that $N(V(P))$ admits the partition



$$\{S_a,S_b,S_c,S_d,S_{ac},S_{ad},S_{bd}\}$$

where $S_a = N(V(P)) \setminus N(b,c,d)$, $S_b = N(V(P)) \setminus N(a,c,d)$, $S_c = N(V(P)) \setminus N(a,b,d)$, $S_d = N(V(P)) \setminus N(a,b,c)$, $S_{a,c} = N(V(P)) \setminus N(b,d)$, $S_{ad} = N(V(P)) \setminus N(b,c)$, $S_{b,d} = N(V(P)) \setminus N(a,c)$.


Then the following observations can be shown with no difficult.

\begin{obse}\label{obse:0}
Every non-trivial component of a ($P_4$, Triangle)-free graph is complete bipartite.  \qed
\end{obse}

\begin{obse}\label{obse:completebipartite}
For any induced $P_4$ of any ($P_4+P_4$, Triangle)-free $G$, say $P$, one has:
\begin{itemize}
  \item [(i)] every non-trivial component of $G[A(V(P))]$ is complete bipartite;
  \item [(ii)] each vertex of $N(V(P))$ does not contact both sides of any non-trivial connected component of $G[A(V(P))]$.  \qed
\end{itemize}
\end{obse}

Then let us recall that WIS can be solved for bipartite graphs in polynomial time
\cite{AhuMagOrl1993,DesHak1970,GroLovSch1988}. In particular let us formalize as lemma the following fact which can be (independently) shown with no difficult.

\begin{lemm}\label{lemm:0}
The WIS problem can be solved for complete bipartite graphs in polynomial time, i.e., in linear time. \qed
\end{lemm}

Then let us introduce a lemma.

\begin{lemm}\label{lemm:1}
Let $G$ be a ($P_4+P_4$, Triangle)-free graph containing an induced $P_4$, say $P$, of vertex set $V(P) = \{a,b,c,d\}$ and edge set $E(P) = \{ab,bc,cd\}$. Then a maximum weight independent set of $G$ containing $\{a,c\}$ (containing $\{b,d\}$, respectively, by symmetry) can be computed in polynomial time.
\end{lemm}

{\bf Proof.} The proof is introduced in Subsection 2.1. \qed \\



Then let us consider the following algorithm. \\

{\bf Algorithm Last}

{\bf Input:} a ($P_4+P_4$, Triangle)-free graph $G$.

{\bf Output:} a maximum weight independent set of $G$.

\

{\em Step 1.}

For each induced $P_4$ of $G$, say $P$, of vertex set $V(P) = \{a,b,c,d\}$ and edge set $E(P) = \{ab,bc,cd\}$ do:

(1.1) compute [by Lemma \ref{lemm:1}] a maximum weight independent set of $G$ containing $\{a,c\}$: denote it as $Q_1$;

(1.2) compute [by Lemma \ref{lemm:1}] a maximum weight independent set of $G$ containing $\{b,d\}$: denote it as $Q_2$;

(1.3) compute [by Lemma \ref{lemm:0}] a maximum weight independent set of $G[\{a,d\} \cup L \cup A(V(P))]$ where $L$ is the set of those vertices in $S_b \cup S_c$ which are isolated in $G[S_b \cup S_c \cup A(V(P))]$: denote it as $Q_3$;

(1.4) select a $best$ weight independent set of $G$ over $\{Q_1,Q_2,Q_3\}$: denote it as $Q(P)$.

{\em Step 2.}

Select a $best$ weight independent set of $G$ over $\{Q(P): P$ is an induced $P_4$ of $G\}$: denote it as $Q_{black}$.

{\em Step 3.}

(3.1) Remove from $G$ all the vertices of $G$ which belong to an induced $P_4$ of $G$: let $G'$ be the graph obtained in this way.

(3.2) Compute [by Lemma \ref{lemm:0}] a maximum weight independent set of $G'$: denote it as $Q_{white}$.

{\em Step 4.}

Select a $best$ weight independent set of $G$ over $\{Q_{black},Q_{white}\}$ and output it. \\

\begin{theo}\label{theo:1}
The WIS problem can be solved for $(P_4+P_4, Triangle)$-free graphs in polynomial time via Algorithm Last.
\end{theo}

{\bf Proof.} First let us show that Algorithm Last can be executed in polynomial time.

As a preliminary let us observe that any (input) graph $G$ contains $O(n^4)$ induced $P_4$'s. Concerning Step 1: steps (1.1)-(1.2) can be executed in polynomial time by Lemma \ref{lemm:1}; step (1.3) can be executed in polynomial time since every connected component of $G[\{a,d\} \cup L \cup A(V(P))]$ is complete bipartite: that follows since by construction $\{a,d\} \cup L$ is an isolated independent set of $G[\{a,d\} \cup L \cup A(V(P))]$ and since by Observation \ref{obse:completebipartite}
each non-trivial component of $G[A(V(P))]$ is complete bipartite; step (1.4) can be executed in constant time; then, by the preliminary observation, Step 1 can be executed in polynomial time. Concerning Step 2: it can be executed in polynomial time by the preliminary observation and since Step 1 can be executed in polynomial time. Concerning Step 3: it can be executed in polynomial time, by the preliminary observation, and since every connected component of $G'$ is complete bipartite by Observation \ref{obse:0}. Concerning Step 4: it can be executed in constant time.

Then let us show that Algorithm Last is correct.


Let $U$ be any maximum (weight) independent set $U$ of $G$: then let us show that Algorithm Last computes $U$ or an $equivalent$ optimal solution.

{\bf Case 1} $U \cap V(P) \neq \emptyset$ for some induced $P_4$ say $P$ of $G$.

Let $V(P) = \{a,b,c,d\}$ and $E(P) = \{ab,bc,cd\}$. Then one has $1 \leq |U \cap V(P)| \leq 2$. 
Then let us consider the following exhaustive subcases.


{\bf Case 1.1} $U \cap V(P) = \{a,c\}$.

Then a maximum weight independent set of $G$ is computed in steps (1.1)-(1.2) with respect to $P$.

{\bf Case 1.2} $U \cap V(P) = \{b,d\}$.

This case can be treated similarly to Case 1.1 by symmetry.

{\bf Case 1.3} $U \cap V(P) = \{a,d\}$.

Then a maximum weight independent set of $G$ is a maximum weight independent set of $G[\{a,d\} \cup S_b \cup S_c \cup A(V(P))]$.
Note that, since $G$ is Triangle-free, $S_b$ and $S_c$ are independent sets.
Then $S_b \cup S_c$ admits a partition, say $\{L,L'\}$, where $L$ is the set of those vertices of $S_b \cup S_c$ which are isolated in $G[S_2 \cup S_3 \cup A(V(P))]$ [as defined above] and $L' = (S_b \cup S_c) \setminus L$. Now: (i) either $U \cap L' = \emptyset$, in which case a maximum weight independent set of $G$ is contained in $\{a,d\} \cup L \cup A(V(P))$, so that it is computed in step (1.3) with respect to $P$; (ii) or $U \cap L' \cap S_b \neq \emptyset$, namely there is a vertex say $b' \in U \cap L' \cap S_b$ with a neighbor say $b'' \in S_c \cup A(V(P))$, so that vertices $a,b,b'b''$ induce a $P_4$ say $P(b)$ of $G$, and then a maximum weight independent set of $G$ is computed in step (1.3) with respect to $P(b)$; (iii) or $U \cap L' \cap S_c \neq \emptyset$, namely there is a vertex say $c' \in U \cap L' \cap S_c$ with a neighbor say $c'' \in S_b \cup A(V(P))$, so that vertices $d,c,c'c''$ induce a $P_4$ say $P(c)$ of $G$, and then a maximum weight independent set of $G$ is computed in step (1.3) with respect to $P(c)$.




{\bf Case 1.4} $U \cap V(P) = \{a\}$.

Note that every maximum weight (thus maximal) independent set of $G[V(G) \setminus N(a)]$ not containing vertices of $\{b,c,d\}$ has to contain some vertex of $S_c$, namely there is a vertex say $c' \in U \cap S_c$, so that vertices $a,b,c,c'$ induce a $P_4$ say $P'$ of $G$, and then a maximum weight independent set of $G$ is computed in step (1.3) with respect to $P'$.




{\bf Case 1.5} $U \cap V(P) = \{b\}$.


Note that every maximum weight (thus maximal) independent set of $G[V(G) \setminus N(b)]$ not containing vertices of $\{a,c,d\}$ has to contain some vertex of $S_d \cup S_{ad}$, namely there is a vertex say $d' \in U \cap (S_d \cup S_{ad})$, so that vertices $b,c,d,d'$ induce a $P_4$ say $P'$ of $G$, and then a maximum weight independent set of $G$ is computed in step (1.3) with respect to $P'$.


{\bf Case 1.6} $U \cap V(P) = \{c\}$.

This case can be treated similarly to Case 1.5 by symmetry.

{\bf Case 1.7} $U \cap V(P) = \{d\}$.

This case can be treated similarly to Case 1.4 by symmetry.

{\bf Case 2} $U \cap V(P) = \emptyset$ for any induced $P_4$ say $P$ of $G$.

Then a maximum weight independent set of $G$ is computed in Step 3.

This completes the proof of the theorem. \qed

\subsection{Proof of Lemma 1}\label{proof of lemma 1}

In this subsection let us introduce the proof of Lemma 1.

Let $G$ be a ($P_4+P_4$, Triangle)-free graph, with vertex weight function $w$, containing an induced $P_4$ say $P$ of vertex set $V(P) = \{a,b,c,d\}$ and edge set $E(P) = \{ab,bc,cd\}$.


Let us show that a maximum weight independent set of $G$ containing $\{a,c\}$ (containing $\{b,d\}$, respectively, by symmetry) can be computed in polynomial time.

A maximum weight independent set of $G$ containing $\{a,c\}$ can be computed by solving WIS for $G[\{a,c\} \cup S_b \cup S_d \cup S_{bd} \cup A(V(P))]$. Then, since vertices of $\{a,c\}$ are isolated in such a graph, the problem can be reduced to graph $G[S_b \cup S_d \cup S_{bd} \cup A(V(P))]$.



Then let us show that WIS can be solved for $G[S_b \cup S_d \cup S_{bd} \cup A(V(P))]$ in polynomial time.


The proof consists of solving a sequence of cases which are more and more difficult/general, each of which is solved by a reduction to the previous solved case, where the basic case is that of complete bipartite graphs. In this sense the proof is not a massive case distinction.


In what follows two main macro-cases are solved, namely, CASE A as the facilitated case and CASE B as the general case.


\subsubsection{CASE A: the facilitated case}


CASE A is the following: 
graph $G$ is such that $V(G)$ admits a partition $\{S, T\}$, where $S$ is an independent set and $T$ induces a $P_4$-free graph, so that by Observation \ref{obse:completebipartite} every non-trivial connected component of $G[T]$ is complete bipartite.

Then let us show that WIS can be solved for $G$ in polynomial time.






Let $v \in S$ and let $H$ be a non-trivial component-set of $G[T]$. Let us say that: $v$ is {\em bi-partial} to $H$ if $v$ is partial to one of the sides of $G[H]$; $v$ is {\em bi-universal} to $H$ if $v$ is universal to one of the two sides of $G[H]$. Then by Observation \ref{obse:completebipartite}, if $v$ contacts $H$, then $v$ is either bi-partial to $H$ or bi-universal to $H$.

Let ${\cal H}$ denote the family of non-trivial component-sets of $G[T]$: then, as recalled above, every member of ${\cal H}$ induces a complete bipartite graph. For any $v \in S$, let ${\cal H}[v]$ be the family of members of ${\cal H}$ contacted by $v$. \\

{\bf CASE A.1} No vertex of $S$ is bi-partial to any member of ${\cal H}$.

Then, according to the above, to our aim each member $H$ of ${\cal H}$,
say of sides $H'$ and $H''$,
can be assumed to be [contracted into] one edge say $h'h''$ by defining the weight of $h'$ and of $h''$ as follows: $w(h') = \sum_{h \in H'}w(h)$ and $w(h'') = \sum_{h \in H''}w(h)$.  \\

{\bf CASE A.1.1} Each vertex of $S$ contacts at most one member of ${\cal H}$.

Then since $G$ is $P_4+P_4$-free, there exists at most one member of ${\cal H}$, i.e., one edge say $h'h''$ of $G[T]$, such that both $h'$ and $h''$ have neighbors in $S$: if such an edge $h'h''$ of $G[T]$ does not exist, then every connected component of $G$ is complete bipartite, and then WIS can be solved for $G$ in polynomial time; if such an edge $h'h''$ of $G[T]$ does exist, then every connected component of both $G[V(G) \setminus N(h')]$ and $G[V(G) \setminus \{h'\}]$ is complete bipartite, and then WIS can be solved for $G$ in polynomial time. \\

{\bf CASE A.1.2} Some vertex of $S$ contacts more than one member of ${\cal H}$.

Let $v' \in S$ be such that $|{\cal H}[v']| \geq |{\cal H}[v]|$ for all $v \in S$. 
Let ${\cal H}_{one}$ denote the family of non-trivial component-sets of $G[T \setminus N(v')]$. Note that each vertex of $S$ contacts at most one member of ${\cal H}_{one}$: in fact, if a vertex $v \in S$ should contact two members of ${\cal H}_{one}$, then by construction and by Observation \ref{obse:completebipartite} vertex $v$ would contact two member of ${\cal H}$, and then by definition of $v'$ there would exist two members of ${\cal H}$ which are contacted by $v'$ and non-contacted by $v$, and then by Observation \ref{obse:completebipartite} an induced $P_4+P_4$ would arise. Then WIS can be solved for $G$ in polynomial time as follows: for $G[V(G) \setminus N(v')]$ one can refer to CASE A.1.1; for $G[V(G) \setminus \{v'\}]$ one can iterate the above argument until the graph is reduced to $G[T]$; for $G[T]$ one can solve WIS in polynomial time since every connected component of $G[T]$ is complete bipartite. \\

{\bf CASE A.2} Some vertex of $S$ is bi-partial to some member of ${\cal H}$. \\

{\bf CASE A.2.1} Each vertex of $S$ is bi-partial to at most one member of ${\cal H}$.

Let $v' \in S$ be bi-partial to one member of ${\cal H}$ and be such that $|{\cal H}[v']| \geq |{\cal H}[v]|$ for all $v \in S$ which are bi-partial to one member of ${\cal H}$: in particular let $H'$ be the member of ${\cal H}$ such that $v'$ is bi-partial to $H'$.

Then let ${\cal Z}$ be the family of non-trivial component-sets $Z$ of $G[(T \setminus H') \setminus N(v')]$ such that there is a vertex of $S$ bi-partial to $Z$. \\

{\bf Claim 1.} {\em ${\cal Z}$ has at most one member.}

{\bf Proof.} By contradiction assume that ${\cal Z}$ has two members, say $Z_1$ and $Z_2$. By definition of ${\cal Z}$, let $v_1,v_2 \in S$ be respectively bi-partial to $Z_1,Z_2$ (actually $v_1$ may coincide to $v_2$; however both $v_1,v_2$ are different to $v$).

Let us observe that: if $v_1$ coincides to $v_2$, then such a vertex contacts both $Z_1$ and $Z_2$; if $v_1$ does not coincide to $v_2$, then to avoid a $P_4+P_4$, either $v_1$ contacts $Z_2$ or $v_2$ contacts $Z_1$. Then, without loss of generality by symmetry, let us assume that $v_1$ contacts [both $Z_1$ and] $Z_2$.

Then by construction and by Observation \ref{obse:completebipartite}, there exist  two members of ${\cal H}$, say $H_1,H_2$, such that $Z_1 \subseteq H_1$ and $Z_2 \subseteq H_2$. By definition of $v'$, one has that $v'$ does not contact $H_1,H_2$: in fact, if $v'$ should contact either $H_1$ or $H_2$, then by construction $v'$ would be bi-partial to it (a contradiction to the assumption of CASE A.2.1, since $v'$ is bi-partial to $H'$). Then, by definition of $v'$, one has that $v'$ contacts at least two members of ${\cal H}$ which are not contacted by $v_1$: then, from one hand the subgraph induced by $v'$ and by such two members contains an induced $P_4$, and from the other hand the subgraph induced by $v_1$ and by $Z_1$ contains an induced $P_4$, i.e., an induced $P_4+P_4$ arises (contradiction). \qed \\

{\bf Claim 2.} {\em WIS can be solved for $G[V(G) \setminus N(v')]$ in polynomial time.}

{\bf Proof.} By Claim 1, ${\cal Z}$ has at most one member. Let us consider only the case in which such a member does exist, say ${\cal Z} = \{Z\}$, since the other case can be treated similarly. Then $H$ be the member of ${\cal H}$ such that $Z \subseteq H$. Note that $H \setminus N(v')$ and $H' \setminus N(v')$ are the only (two) non-trivial component-sets of $G[T \setminus N(v')]$ to which any vertex of $S$ may be bi-partial. Furthermore by Observation \ref{obse:completebipartite}, for any $h \in H$ (for any $h' \in H'$, respectively), $h$ is universal to one side of $H$ ($h'$ is universal to one side of $H'$, respectively).

For any maximum (weight) independent set $U$ of $G$ one of the following cases occurs: (i) $U \cap H = \emptyset$ and $U \cap H' = \emptyset$, (ii) $U \cap H = \emptyset$ and $U \cap H' \neq \emptyset$, (iii) $U \cap H \neq \emptyset$ and $U \cap H' = \emptyset$, (iv) $U \cap H \neq \emptyset$ and $U \cap H' \neq \emptyset$.

Then WIS can be solved for $G[V(G) \setminus N(v')]$ as follows.


In case (i): by solving WIS for $G[(V(G) \setminus N(v')) \setminus (H \cup H')]$, which enjoys CASE A.1 by the above. In case (ii): by solving WIS for $G[(V(G) \setminus N(v')) \setminus N(h')]$, for all $h' \in H'$, which enjoys CASE A.1 by the above. In case (iii): by solving WIS for $G[(V(G) \setminus N(v')) \setminus N(h)]$, for all $h \in H$, which enjoys CASE A.1 by the above. In case (iv): by solving WIS for $G[(V(G) \setminus N(v')) \setminus N(h,h')]$, for all $(h,h') \in H \times H'$, which enjoys CASE A.1 by the above.

Then WIS can be solved for $G[V(G) \setminus N(v')]$ in polynomial time by referring to CASE A.1.  \qed \\

Then WIS can be solved for $G$ in polynomial time as follows: for $G[V(G) \setminus N(v')]$ one can refer to Claim 2, that is, finally to CASE A.1; for $G[V(G) \setminus \{v'\}]$ one can iterate the above argument until the graph is reduced to $G[T]$; for $G[T]$ one can solve WIS in polynomial time since every connected component of $G[T]$ is complete bipartite. \\

{\bf CASE A.2.2} Some vertex of $S$ is bi-partial to more than one member of ${\cal H}$.

Let us define an order $' < '$ on $S$: let us say that, for any $u,v \in S$, $u < v$ if $v$ is bi-partial to two non-trivial component-sets of $G[T \setminus N(u)]$. \\

{\bf Claim 3.} The ordered set ($S, <$) admits a maximal element, i.e., there exists $v^* \in S$ such that no vertex of $S$ is bi-partial to two non-trivial component-sets of $G[T \setminus N(v^*)]$. In particular, WIS can be solved for $G[V(G) \setminus N(v^*)]$ in polynomial time. \\

{\bf Proof.} As a preliminary let us introduce the following observation. Let $u,v \in S$ and assume $u < v$, that is, $v$ be bi-partial to two non-trivial component-sets, say $Z_1,Z_2$, of $G[T \setminus N(u)]$: then by construction and by Observation \ref{obse:completebipartite} there exist two members of ${\cal H}$, say $H_1,H_2$, such that $Z_1 \subseteq H_1$ and $Z_2 \subseteq H_2$.

Then let us prove the following facts. \\

{\em Fact 1.} Let $u,v \in S$ and assume $u < v$, that is, $v$ be bi-partial to two non-trivial component-sets, say $Z_1,Z_2$, of $G[T \setminus N(u)]$; then let $H_1,H_2$ be the two members of ${\cal H}$ such that $Z_1 \subseteq H_1$ and $Z_2 \subseteq H_2$. Then: if $u$ contacts $H_1$ (contacts $H_2$, respectively), then $N_{H_1}(u) \subset N_{H_1}(v)$ (then $N_{H_2}(u) \subset N_{H_2}(v)$, respectively).

{\em Proof of Fact 1.} By contradiction assume that $u$ contacts $H_1$ and that $N_{H_1}(u) \not \subset N_{H_1}(v)$ (i.e., $u$ is adjacent to a vertex of $H_1 \setminus Z_1$ non-adjacent to $v$). Then, by Observation \ref{obse:completebipartite} and since $v$ is bi-partial to $Z_1$, one has that (considering that $u$ may contact $H_1$ either in the same side as $v$ or in the other side):
from one hand the subgraph induced by $u$ and $H_1$ contains an induced $P_4$ not contacted by $v$, and from the other hand the subgraph induced by $v$ and by $Z_2$ contains an induced $P_4$, i.e., an induced $P_4+P_4$ arises (contradiction). The same holds for $H_2$ instead of $H_1$ by symmetry.  \qed \\

{\em Fact 2.} Let $u,v \in S$ and assume $u < v$. Then $v \not < u$.

{\em Proof of Fact 2.} By assumption let $v$ be bi-partial to two non-trivial component-sets, say $Z_1,Z_2$, of $G[T \setminus N(u)]$. Then let $H_1,H_2$ be the two members of ${\cal H}$ such that $Z_1 \subseteq H_1$ and $Z_2 \subseteq H_2$. By contradiction assume that $v < u$. Then let
$u$ be bi-partial to two non-trivial component-sets, say $Z_3,Z_4$, of $G[T \setminus N(v)]$. Then let $H_3,H_4$ be the two members of $K$ such that $Z_3 \subseteq H_3$ and $Z_4 \subseteq H_4$.

Then by Fact 1 one that $H_3 \neq H_1,H_2$ and that $H_4 \neq H_1,H_2$. Then, from one hand the subgraph induced by $v$ and $Z_1$ contains an induced $P_4$, and from the other hand the subgraph induced by $u$ and $Z_3$ contains an induced $P_4$, i.e., an induced $P_4+P_4$ arises (contradiction).  \qed \\

Now let $v_1,v_2,\ldots,v_p \in S$, for some natural $p \geq 3$, and assume $v_1 < v_2 < \ldots < v_p$. Then $v_j$ is bi-partial to two non-trivial component-sets, say $Z_1(j),Z_2(j)$, of $G[T \setminus N(v_{j-1})]$ for $j \in \{2, \ldots, p\}$. Then let $H_1(j),H_2(j)$ be the two members of ${\cal H}$ such that $Z_1(j) \subseteq H_1(j)$ and $Z_2(j) \subseteq H_2(j)$ for $j \in \{2, \ldots, p\}$. \\

{\em Fact 3.} $v_p$ contacts $Z_1(2),Z_2(2)$.

{\em Proof of Fact 3.} First let us show that $v_p$ contacts $Z_1(p-1),Z_2(p-1)$. Let us show that $v_p$ contacts $Z_1(p-1)$. If either $H_1(p-1) = H_1(p)$ or $H_1(p-1) = H_2(p)$, say $H_1(p-1) = H_1(p)$ (without loss of generality by symmetry), then by construction $N_{H_1(p)}(v_{p-1}) \subseteq  H_1(p) \setminus Z_1(p)$, that is $Z_1({p-1}) \subseteq Z_1(p)$, that is $v_p$ contacts $Z_1(p-1)$. If $H_1(p-1) \neq H_1(p), H_2(p)$, then $v_p$ contacts $Z_1(p-1)$, since otherwise a $P_4+P_4$ arises (one $P_4$ is contained in the subgraph induced by $v_{p-1}$ and $Z_1({p-1})$, one $P_4$ is contained in the subgraph induced by $v_{p}$, $Z_1(p)$, $Z_2(p)$). Then $v_p$ contacts $Z_1(p-1)$. The same holds for $Z_2(p-1)$ by symmetry.

Then let us show that for $3 \leq j \leq p-1$, if $v_p$ contacts $Z_1(j),Z_2(j)$, then $v_p$ contacts $Z_1(j-1),Z_2(j-1)$. Let us show that if $v_p$ contacts $Z_1(j),Z_2(j)$, then $v_p$ contacts $Z_1(j-1)$. If either $H_1(j-1) = H_1(j)$ or $H_1(j-1) = H_2(j)$, say $H_1(j-1) = H_1(j)$ (without loss of generality by symmetry), then by construction $N_{H_1(j)}(v_{j-1}) \subseteq  H_1(j) \setminus Z_1(j)$, that is $Z_1({j-1}) \subseteq Z_1(j)$, that is $v_p$ contacts $Z_1(j-1)$. If $H_1(j-1) \neq H_1(j), H_2(j)$, then $v_p$ contacts $Z_1(j-1)$, since otherwise a $P_4+P_4$ arises (one $P_4$ is contained in the subgraph induced by $v_{p-1}$ and $Z_1({j-1})$, one $P_4$ is contained in the subgraph induced by $v_{p}$, $Z_1(j)$, $Z_2(j)$). Then $v_p$ contacts $Z_1(j-1)$. The same holds for $Z_2(j-1)$ by symmetry.

Then Fact 3 is proved. \qed \\

{\em Fact 4.} $v_p \not < v_1$.

{\em Proof of Fact 4.} By contradiction assume $v_p < v_1$. Then $v_1$ is bi-partial to two non-trivial component-sets, say $Z_1,Z_2$, of $G[T \setminus N(v_p)]$. Then let $H_1,H_2$ be the two members of ${\cal H}$ such that $Z_1 \subseteq H_1$ and $Z_2 \subseteq H_2$. Let us recall that $v_p$ contacts $Z_1(2),Z_2(2)$ by Fact 3. If either $H_1 = H_1(2)$ or $H_1 = H_2(2)$, say $H_1 = H_1(2)$ (without loss of generality by symmetry), then by construction $N_{H_1(2)}(v_{1}) \subseteq  H_1(2) \setminus Z_1(2)$, that is $Z_1 \subseteq Z_1(2)$, that is $v_p$ contacts $Z_1$ (contradiction).
If $H_1 \neq H_1(2), H_2(2)$, then $v_p$ contacts $Z_1$ (contradiction), since otherwise a $P_4+P_4$ arises (one $P_4$ is contained in the subgraph induced by $v_{1}$ and $Z_1$, one $P_4$ is contained in the subgraph induced by $v_{p}$, $Z_1(2)$, $Z_2(2)$). \qed \\


Let us conclude the proof of Claim 3. By Facts 2 and 4, there are no vertices $u_1,u_2,\ldots,u_k \in S$ (for $k \geq 2$) such that $u_1 < u_2 < \ldots < u_k < u_1$, i.e., there is no {\em cycle} in the ordered set $(S, <)$. Then ($S, <$) admits a maximal element, i.e., there exists $v^* \in S$ such that no vertex of $S$ is bi-partial to two non-trivial component-sets of $G[T \setminus N(v^*)]$.

The above fact can be seen (more formally) by defining a {\em directed graph}, namely $D=(S,E(S))$, such that for any $u,v \in S$ there is directed edge $(u,v)$ if and only if $u < v$; then by the above $D$ is acyclic; then it is well-known [and not difficult to check] that $D$ contains at least one vertex with zero out-degree.

In particular, WIS can be solved for $G[V(G) \setminus N(v^*)]$ in polynomial time, since $G[V(G) \setminus N(v^*)]$ enjoys CASE A.2.1. \qed \\


Then WIS can be solved for $G$ in polynomial time as follows: compute a maximal element of ($S, <$), say $v^*$, and solve WIS for $G[V(G) \setminus N(v^*)]$ by Claim 3, that is, by finally referring to CASE A.2.1; iterate this procedure for $G[V(G) \setminus \{v^*\}]$ until the graph is reduced to $G[T]$; solve WIS for $G[T]$ in polynomial time since every connected component of $G[T]$ is complete bipartite.

This completes the solution for CASE A.

\subsection{CASE B: the general case.}


Let us show that WIS can be solved for $G[S_b \cup S_d \cup S_{bd} \cup A(V(P))]$ in polynomial time. Let us recall that $S_b \cup S_{bd}$ and  $S_d \cup S_{bd}$ are independent sets and that every non-trivial component of $G[A(V(P))]$ is complete bipartite.

For brevity let us write $T = A(V(P))$.







Let $v \in S_b \cup S_d \cup S_{bd}$ and let $H$ be a non-trivial component-set of $G[T]$. Let us say that: $v$ is {\em bi-partial} to $H$ if $v$ is partial to one of the sides of $G[H]$; $v$ is {\em bi-universal} to $H$ if $v$ is universal to one of the two sides of $G[H]$. Then by Observation \ref{obse:completebipartite}, if $v$ contact $H$, then $v$ is either bi-partial to $H$ or bi-universal to $H$.



For any maximum (weight) independent set $U$ of $G[S_b \cup S_d \cup S_{bd} \cup T]$ one of the following cases occurs: (i) $U \cap S_b = \emptyset$ and $U \cap S_d = \emptyset$, (ii) $U \cap S_b = \emptyset$ and $U \cap S_d \neq \emptyset$, (iii) $U \cap S_b \neq \emptyset$ and $U \cap S_d = \emptyset$, (iv) or $U \cap S_b \neq \emptyset$ and $U \cap S_d \neq \emptyset$.

Then WIS can be solved for $G[S_b \cup S_d \cup S_{bd} \cup T]$ as follows.

In case (i): by solving WIS for $G[S_{bd} \cup T]$, in polynomial time, since it enjoys CASE A. In case (ii): by solving WIS for $G[S_d \cup S_{bd} \cup T]$, in polynomial time, since it enjoys CASE A. In case (iii): by solving WIS for $G[S_b \cup S_{bd} \cup T]$, in polynomial time, since it enjoys CASE A. In case (iv): by solving WIS for $G[(S_b \cup S_d \cup S_{bd} \cup T) \setminus N(s_b,s_d)]$ for all non-adjacent pair of vertices $(s_b,s_d) \in S_b \times S_d$.

Then $-$ to show that WIS can be solved for $G[S_b \cup S_d \cup S_{bd} \cup T]$ in polynomial time $-$ it remains to show that WIS can be solved for $G[(S_b \cup S_d \cup S_{bd} \cup T) \setminus N(s_b,s_d)]$ in polynomial time for all non-adjacent pair of vertices $(s_b,s_d) \in S_b \times S_d$.

Then let us write $G' = G[(S_b \cup S_d \cup S_{bd} \cup T) \setminus N(s_b,s_d)]$ for any fixed $(s_b,s_d) \in S_b \times S_d$.

Then let us write $S'_i = S_i \setminus N(s_b,s_d)$ for $i = b,d,bd$, and $T' = T \setminus N(s_b,s_d)$: then $G' = G[\{s_b,s_d\} \cup S'_b \cup S'_d \cup S'_{bd} \cup T']$.

Let ${\cal H}'_{all}$ denote the family of [all, i.e., trivial or non-trivial] component-sets of $G[T']$. For any $v \in S'_b \cup S'_d$, let ${\cal H}'_{all}[v]$ be the family of members of ${\cal H}'_{all}$ contacted by $v$.

Let $v' \in S'_b \cup S'_d$ such that: (j) $|{\cal H}'_{all}[v']| \geq |{\cal H}'_{all}[v]|$ for all $v \in S'_b \cup S'_d$, and (jj) $N_{T'}(v') \not \subset N_{T'}(v)$ for all $v \in (S'_b \cup S'_d) \setminus \{v'\}$.

Let us assume that $v' \in S'_b$ without loss of generality by symmetry.

Let us show that WIS can be solved for $G'[V(G') \setminus N(v')]$ in polynomial time.

Let us write $G'' = G'[V(G') \setminus N(v')]$. 

Then let us write $S''_i = S'_i \setminus N(v')$ for $i = b,d,bd$, and $T'' = T' \setminus N(v')$: then $G'' = G[\{s_b,s_d,v'\} \cup S''_b \cup S''_d \cup S''_bd \cup T'']$.

Let ${\cal H}'$ denote the family of non-trivial component-sets of $G[T']$. \\


{\bf CASE B.1} No vertex of $S''_d$ is bi-partial to any member of ${\cal H}'$.


Then let us prove the following facts. \\

{\em Fact 1.} Each vertex of $S''_d$ contacts no component-set of $G[T']$ not contacted by $v'$.

{\em Proof of Fact 1.} By contradiction assume that a vertex $v \in S''_d$ contacts a component-set say $H$ of $G[T']$ not contacted by $v'$, i.e., $v$ is adjacent to a vertex $h \in H$ with $H$ not contacted by $v'$. Then by definition of $v'$, there is a vertex $h' \in T' \setminus H$ which is adjacent to $v'$ and non-adjacent to $v$.
Then $s_b,b,v,h$ and $s_d,d,v',h'$ induce a $P_4+P_4$ (contradiction). \qed \\

{\em Fact 2.} Each vertex of $S''_d$ has neighbors, which are non-neighbors of $v'$, in at most one component-set of $G[T']$.

{\em Proof of Fact 2.} By contradiction assume that a vertex $v \in S''_d$ has neighbors say $h_1,h_2$, which are non-neighbors of $v'$, in respectively two component-sets say $H_1,H_2$ of $G[T']$. By Fact 1, $v'$ contacts $H_1,H_2$, i.e., $v'$ has neighbors say $h'_1,h'_2$ in respectively $H_1,H_2$. Then $h'_1$ is adjacent to $h_1$, and $h'_2$ is adjacent to $h_2$, since otherwise a $P_4+P_4$ arises. Then $s_b,b,v',h'_1$ and $s_d,d,v,h_2$ induce a $P_4+P_4$ (contradiction). \qed \\

{\em Fact 3.} $G[S''_d \cup T'']$ is $P_4$-free.

{\em Proof of Fact 3.} By contradiction assume that $G[S''_d \cup T'']$ contains an induced $P_4$, say $P^*$, of vertex-set $V(P^*)$. Then: from one hand $|V(P^*) \cap S''_d| \geq 1$, since $G[T'']$ is $P_4$-free; from the other hand $|V(P^*) \cap S''_d| \leq 2$, since $S''_d$ is an independent set.

The occurrence $|V(P^*) \cap S''_d| = 1$ is not possible by Observation \ref{obse:completebipartite} and by Fact 2 with respect to the vertex of $V(P^*) \cap S''_d$.

The occurrence $|V(P^*) \cap S''_d| = 2$ is not possible as shown in the following sub-occurrences.

Assume that $V(P^*) = \{u,x,v,y\}$, with $u,v \in S''_d$ and $x,y \in T''$, with edges $ux,xv,vy$. Then, since $v$ is adjacent to both $x$ and $y$, by Fact 2 vertices $x,y$ belong to the same component-set of $G[T']$. But this contradicts the assumption of Case 1 with respect to $u$.

Assume that $V(P^*) = \{u,x,y,v\}$, with $u,v \in S''_d$ and $x,y \in T''$, with edges $ux,xy,vy$. Then, since vertices $x,y$ are adjacent, vertices $x,y$ belong to the same component-set of $G[T'']$ (= $G[T' \setminus N(v')]$), say $Z$, and to different sides of $Z$ respectively. Then let $H$ be the component-set of $G[T']$ such that $Z \subseteq H$. By Fact 1, vertex $v'$ contacts $H \setminus Z$, i.e., vertex $v'$ contacts one side of $G[H \setminus Z]$: without loss of generality by symmetry say $v'$ contacts the side of $G[H \setminus Z]$ corresponding to the side of $Z$ contacted by $u$. Then for any neighbors of $v'$ in $H \setminus Z$, say $h$, one has that $u$ is adjacent to $h$, since otherwise $s_b,b,v',h$ and $s_d,d,u,x$ induce a $P_4+P_4$. That is one has $N_H(v') \subset N_H(u)$. Then, since $N_{T'}(v') \not \subset N_{T'}(u)$ (by definition of $v'$), there is a vertex $h' \in T' \setminus H$ such that $h'$ is adjacent to $v'$ and non-adjacent to $u$. Then $s_b,b,v',h'$ and $s_d,d,u,x$ induce a $P_4+P_4$, a contradiction.  \qed \\

Then WIS can be solved for $G''$ in polynomial time, since $\{s_b,s_d,v'\} \cup S''_b \cup S''_{bd}$ is an independent set and since $G[S''_d \cup T'']$ is $P_4$-free by Fact 3, that is since $G''$ enjoys CASE A. \\

{\bf CASE B.2} Some vertex of $S''_d$ is bi-partial to some member of ${\cal H}'$. \\

{\bf CASE B.2.1} Each vertex of $S''_d$ is bi-partial to at most one member of ${\cal H}'$.

This case can be treated similarly to CASE A.2.1, in order to conclude that WIS can be solved for $G''$ in polynomial time by referring to CASE B.1. \\










{\bf CASE B.2.2} Some vertex of $S''_d$ is bi-partial to more than one member of ${\cal H}'$.

This case can be treated similarly to CASE A.2.2, in order to conclude that WIS can be solved for $G''$ in polynomial time by referring to CASE B.2.1. \\







Summarizing CASE B.1 and CASE B.2 one has that: WIS can be solved for $G''$ in polynomial time.

Then WIS can be solved for $G'$ $(= G[\{s_b,s_d\} \cup S'_b \cup S'_d \cup S'_{bd} \cup T'])$ in polynomial time as follows: for $G'[V(G') \setminus N(v')]$  ($= G''$) one can proceed as above; for $G'[V(G') \setminus \{v'\}]$ one can iterate the above argument until the graph is reduced to $G'[\{s_b,s_d\} \cup S'_{bd} \cup T']$; for $G'[\{s_b,s_d\} \cup S'_{bd} \cup T']$ one can refer to CASE A. Then, as remarked above, this implies that WIS can be solved for $G$ in polynomial time.

This completes the solution for CASE B.

\section{Concluding remarks}

Let us list some possible concluding remarks.

1. In \cite{Olari1988}, it is shown that every connected Paw-free graph is either Triangle-free or complete multipartite [a $Paw$ has vertices $a,b,c,d$, and edges $ab,ac,ad,bc$]. This result and Theorem \ref{theo:1} directly imply that the WIS problem can be solved for ($P_4+P_4$, Paw)-free graphs in polynomial time. Furthermore in \cite{Olari1990}, it is shown that if a prime graph contains a Triangle then it contains a House, or a Bull, or a Double-Gem [a $House$ has vertices $a,b,c,d,e$, and edges $ab,ac,bc,be,cd,de$; a $Bull$ has vertices $a,b,c,d,e$, and edges $ab,ac,bc,be,cd$; a $Double$-$Gem$ has vertices $a,b,c,d,e,f$, and edges $ac,ad,ae,bd,be,bf,cd,de,ef$]. This result and Theorem \ref{theo:1}, by well known results on prime graphs (see e.g. \cite{LozMil}),  imply that the WIS problem can be solved for ($P_4+P_4$, House, Bull, Double-Gem)-free graphs in polynomial time.

2. The proof of Theorem \ref{theo:1} is based on the anti-neighborhood approach by finally reducing the problem to instances of complete bipartite graphs for which the problem can be solved in linear time. Then the time bound of Theorem \ref{theo:1}, i.e., of Algorithm Last, may be estimated as $O(n^{15})$ time.

Then one can derive the following result, which is similar to the corresponding results obtained for ($P_7$,Triangle)-free graphs \cite{BraMosP7K3} and for ($S_{1,2,4}$,Triangle)-free graphs \cite{BraMosS124K3}, and which seems to be harmonic [together with such results] with respect to the result of Pr\"omel et al. \cite{ProSchSte2002} showing that with ``high probability'' removing a single vertex in a Triangle-free graph leads to a bipartite graph.

\begin{theo}
For every ($P_4 + P_4$,Triangle)-free graph $G$ there is a family ${\cal S}$ of subsets of $V(G)$ inducing (complete) bipartite subgraphs of $G$, which contains polynomially many members and can be computed in polynomial time, such that every maximal independent set of $G$ is contained in some member of ${\cal S}$. \qed
\end{theo}

An outline of the proof: concerning Lemma \ref{lemm:1} the above result can be derived with no difficult (for every maximal independent sets of $G$ containing vertices $a,c$) by the proof scheme; concerning Theorem \ref{theo:1}, in particular concerning Algorithm Last, the above result can be derived by considering the following alternative step (1.3) of Algorithm Last [in fact Algorithm Last is given in a version which directly aims to solve the WIS problem] according to Case 1.3 of the proof of Theorem \ref{theo:1}:

(1.3) compute [by Lemma \ref{lemm:0}] a maximum weight independent set of $G[\{a,d\} \cup L \cup A(V(P))]$ where $L$ is the set of those vertices in $S_b \cup S_c$ which are isolated in $G[S_b \cup S_c \cup A(V(P))]$: denote it as $Q'_3$; compute [by Lemma \ref{lemm:1}] a maximum weight independent set of $G$ containing $\{a,b'\}$ [which are vertices of an induced $P_4$] and containing $\{d\}$, for every $b' \in L' \cap S_{b}$, where $L' = (S_b \cup S_c) \setminus L$: denote it as denote it as $Q''_3$; compute [by Lemma \ref{lemm:1}] a maximum weight independent set of $G$ containing $\{d,c'\}$ [which are vertices of an induced $P_4$] and containing $\{a\}$, for every $c' \in L' \cap S_{c}$, where $L' = (S_b \cup S_c) \setminus L$: denote it as denote it as $Q'''_3$; finally select a $best$ maximum weight independent set over $\{Q'_3,Q''_3,Q'''_3\}$: denote it as $Q_3$.


3. Finally let us point out the following possible open problem.

{\bf Open Problem.} What is the complexity of (W)IS for $P_4+P_4$-free graphs? \\


{\bf Acknowledgements.} Please I would like to witness that I just try to pray a lot and am not able to do anything without that - ad laudem Domini. \\

\begin{footnotesize}
\renewcommand{\baselinestretch}{0.4}

\end{footnotesize}


\begin{thebibliography}{99}

\bibitem{AhuMagOrl1993}
  R.K. Ahuja, T.L. Magnanti, J.B. Orlin,
  Network Flows,
  Prentice Hall 1993


\bibitem{Aleks1983}
  V.E. Alekseev,
  On the local restriction effect on the complexity of finding the graph independence number,
  {\sl Combinatorial-algebraic Methods in Applied Mathematics}, Gorkiy University Press,
  Gorkiy (1983) 3-13 (in Russian)

\bibitem{Aleks1991}
  V.E. Alekseev,
  On the number of maximal independent sets in graphs from hereditary classes,
  {\sl Combinatorial-algebraic Methods in Discrete Optimization}, Gorkiy University Press,
  Gorkiy (1991) 5-8 (in Russian)

\bibitem{Aleks2004/1}
  V.E. Alekseev,
  A polynomial algorithm for finding largest independent sets in fork-free graphs,
  {\sl Discrete Analysis and Operations Research} Ser. 1, 6 (1999) 3-19 (in Russian), {\sl Discrete Applied Mathematics} 135 (2004) 3-16

\bibitem{Aleks2004/2}
  V.E. Alekseev,
  On easy and hard hereditary classes of graphs with respect to the independent set problem,
  {\sl Discrete Applied Mathematics} 132 (2004) 17-26





\bibitem{BraLeSpi1999}
    A. Brandst\"adt, V.B. Le, J.P. Spinrad,
    Graph Classes: A Survey,
    {\sl SIAM Monographs on Discrete Math. Appl., Vol.} 3,
    SIAM, Philadelphia (1999)





\bibitem{BraMosP7K3}
	A. Brandst\"adt, R. Mosca,
    Stable Sets in ($P_7$, Triangle)-Free Graphs,
    {\sl Discrete Applied Mathematics}, 236 (2018) 57–65

\bibitem{BraMoslClaw}
    A. Brandst\"adt, R. Mosca,
    Maximum weight independent set for $\ell $claw-free graphs in polynomial time, {\em Discrete Applied Mathematics} 237 (2018) 57–64 (2018)

\bibitem{BraMosS124K3}
	A. Brandst\"adt, R. Mosca,
    Maximum Weight Independent Sets for ($S_{1,2,4}$, Triangle)-Free Graphs in Polynomial Time. CoRR,abs/1806.09472, 2018.




\bibitem{CorLerSte2004}
    D.G. Corneil, Y. Perl, L.K. Stewart,
    A linear recognition algortihm for cographs,
    {\sl SIAM J. Computing}
    14 (1985) 926-934

\bibitem{DesHak1970}
    J.F. Desler, S.L. Hakimi,
    On finding a maximum stable set of a graph,
    {\sl Proc. 4th Annual Princeton Conf. on Information Science and Systems},
    Princeton, NJ, 1970




\bibitem{FaeOriSta2011}
    Y. Faenza, G. Oriolo, G. Stauffer,
    An algorithmic decomposition of claw-free graphs leading to an O($n^3$)-algorithm for the weighted independent set problem,
    SODA 2011: 630-646

\bibitem{Farber1989}
    M. Farber,
    On diameters and radii of bridged graphs,
    {\sl Discrete Mathematics} 73 (1989) 249-260

\bibitem{Farber1993}
    M. Farber, M. Hujter, Zs. Tuza,
    An upper bound on the number of cliques in a graph,
    {\sl Networks} 23 (1993) 75-83


\bibitem{FriHedJac}
    G.H. Fricke, S.T. Hedetniemi, D.P. Jacobs,
    Indpependence and irredundance in $k$-regular graphs,
    {\sl Ars Combinatoria} 49 (1998) 271-279


\bibitem{GareyJohnStock1976}
    M.R. Garey, D.S. Johnson, L. Stockmeyer,
    Some simplified NP-complete graph problems,
    {\sl Theoretical Computer Science} 1 (1976) 237-267

\bibitem{GareyJohn1977}
    M.R. Garey, D.S. Johnson,
    The rectilinear Steiner tree problem is NP-complete,
    {\sl SIAM J. Applied Mathematics} 32 (1977) 826-834

\bibitem{GareyJohn1979}
    M.R. Garey, D.S. Johnson,
    Computers and Intractability: A Guide to the Theory of NP-completness,
    Freeman, San Francisco, CA (1979)


\bibitem{GroLovSch1984}
    M. Gr\"otschel, L. Lov\'asz, A. Schrijver,
    Polynomial algorithms for perfect graphs,
    {\sl Annals of Discrete Mathematics} 21 (1984) 325-356

\bibitem{GroLovSch1988}
    M. Gr\"otschel, L. Lov\'asz, A. Schrijver,
    Geometric algorithms and combinatorial optimization,
    Springer, Berlin, 1988

\bibitem{P6}
    Andrzej Grzesik, Tereza Klimosova, Marcin Pilipczuk, and Michal
    Pilipczuk,
    Polynomial-time algorithm for Maximum Weight Independent
    Set on P6-free graphs. In Timothy M. Chan, editor, Proceedings of the 18
    Thirtieth Annual ACM-SIAM Symposium on Discrete Algorithms, SODA
    2019, San Diego, California, USA, January 6-9, 2019, pages 1257–1271.
    SIAM, 2019.







\bibitem{P5}
    D. Lokshantov, M. Vatshelle, Y. Villanger,
    Independent Sets in $P_5$-free Graphs in Polynomial Time,
    http://www.ii.uib.no/~martinv/Papers/ISinP5free.pdf
    In Chandra Chekuri, editor, Proceedings of the Twenty-Fifth Annual ACM-SIAM Symposium on Discrete Algorithms, SODA 2014, Portland, Oregon, USA, January 5-7, 2014, pages 570–581. SIAM, 2014.

\bibitem{Lozin2017}
    V.V. Lozin,
    From matchings to independent sets,
    {\sl Discrete Applied Mathematics}, 231 (2017) 4-14


\bibitem{LozMil}
    V.V. Lozin, M. Milani\v c,
    A polynomial algorithm to find an independent set of maximum
    weight in a fork-free graph,
    {\sl J. Discrete Algorithms} 6 (2008) 595-604

	
\bibitem{LozMos2005}
    V.V. Lozin, R. Mosca,
    Independent sets in extensions of $2K_2$-free graphs,
    {\sl Discrete Applied Mathematics} 146 (2005) 74-80


\bibitem{2P3}
    V.V. Lozin, R. Mosca,
    Maximum regular subgraphs in $2P_3$-free graphs,
    {\sl Theoretical Computer Science} 460 (2012) 26-33

\bibitem{MafPas2016}
    F. Maffray and L. Pastor,
    Maximum weight stable set in ($P_7$, bull)-free graphs and ($S_{1,2,3}$, bull)-free graphs,
    {\sl Discrete Mathematics} 341 (2018) 1449-1458.




\bibitem{Minty1980}
    G.J. Minty,
    On maximal independent sets of vertices in claw-free graphs,
    {\sl J. Combinatorial Theory, Series B} 28 (1980) 284-304




\bibitem{Murphy1992}
    O.J. Murphy,
    Computing independent sets in graphs with large girth,
    {\sl Discrete Applied Mathematics} 35 (1992) 167-170

\bibitem{NakTam2001}
    D. Nakamura, A. Tamura,
    A revision of Minty's algorithm for finding a maximum weight stable set in a claw-free graph,
    {\sl J. Operations Research Society of Japan} 44 (2001) 194-204


\bibitem{NobSas2011}
    P. Nobili, A. Sassano,
    An O($n^2log(n)$) algorithm for the weighted stable set problem in claw-free graphs, CoRR,abs/1501.05775v7, 2019


\bibitem{Olari1988}
    S. Olariu,
    Paw-free graphs,
    {\sl Infomation Processing Letters} 28 (1988) 53-54

\bibitem{Olari1990}
    S. Olariu,
    On the closure of Triangle-free graphs under substitution,
    {\sl Infomation Processing Letters} 34 (1990) 97-101

\bibitem{MSTT2020}
    M. Pilipczuk, N.L.D. Sintiari, S. Thomassé, N. Trotignon,
    (Theta, triangle)-free and (even hole, K4)-free graphs. Part 2 : bounds on treewidth, CoRR,abs/2001.01607, 2020


\bibitem{Polja1974}
    S. Poljak,
    A note on stable sets and colorings of graphs,
    {\sl Commun. Math. Univ. Carolinae} 15 (1974) 307-309

\bibitem{Prisner1995}
    E. Prisner,
    Graphs with few cliques,
    Graph Theory,
    Combinatorics and Algorithms, vol. 1, 2
    (Kalamazoo, MI, 1992),
    Wiley-Interscience Publishers, Wiley, New York, 1995, 945-956

\bibitem{ProSchSte2002}
    H.-J. Pr\"omel, T. Schickinger, A. Steger,
    A note on Triangle-free and bipartite graphs,
    {\sl Discrete Math.} 257 (2002) 531-540




\bibitem{Sbihi1980}
    N. Sbihi,
    Algorithme de recherche d'un independent de cardinalit\'e maximum dans un graphe sans \'etoile,
    {\sl Discrete Mathematics} 29 (1980) 53-76

\bibitem{TsuIdeAriShi1977}
    S. Tsukiyama, M. Ide, H. Ariyoshi, I. Shirakawa,
    A new algorithm for generating all maximal independent sets,
    {\sl SIAM J. on Computing} 6 (1977) 505-517

\end{thebibliography}
\end{document}